\documentclass[12pt,a4paper]{article}
\usepackage{amsmath,amssymb,mathrsfs, url,cite,ifpdf,slashed,multirow,tabularx,type1cm}
\ifpdf        
  \usepackage{graphicx}
\else                          
  \usepackage[dvipdfmx]{graphicx}      
\fi
\allowdisplaybreaks
\usepackage{caption}
\usepackage[colorlinks=true,linkcolor=blue,citecolor=blue,urlcolor=magenta]{hyperref}

\begin{document}


\newcommand{\TeV}{\,{\rm TeV}}
\newcommand{\GeV}{\,{\rm GeV}}
\newcommand{\MeV}{\,{\rm MeV}}
\newcommand{\keV}{\,{\rm keV}}
\newcommand{\invfb}{\,{\rm fb^{-1}}}
\newcommand{\invpb}{\,{\rm pb^{-1}}}
\newcommand{\Slash}[1]{{\ooalign{\hfil \hspace*{-5pt}~#1\hfil\crcr\raise.167ex\hbox{/}}}}
\def\be{\begin{equation}}
\def\ee{\end{equation}}
\def\beq{\begin{eqnarray}}
\def\eeq{\end{eqnarray}}
\def\fr{\frac}
\def\({\left(}
\def\){\right)}
\def\<{\langle}
\def\>{\rangle}
\newcommand{\non}{\nonumber \\ }
\newcommand{\matl}{\left( \begin{array}}
\newcommand{\matr}{\end{array} \right)}
\renewcommand{\sb}{\sin\beta}
\newcommand{\cb}{\cos\beta}
\newcommand{\tb}{\tan\beta}
\newcommand{\eq}[1]{Eq.~(\ref{#1})}

\def\o{\over}
\newcommand{\gsim}{ \mathop{}_{\textstyle \sim}^{\textstyle >} }
\newcommand{\lsim}{ \mathop{}_{\textstyle \sim}^{\textstyle <} }
\newcommand{\vev}[1]{ \left\langle {#1} \right\rangle }
\newcommand{\bra}[1]{ \langle {#1} | }
\newcommand{\ket}[1]{ | {#1} \rangle }
\newcommand{\EV}{ \text{eV} }
\newcommand{\KEV}{ \text{keV} }
\newcommand{\MEV}{ \text{MeV} }
\newcommand{\GEV}{ \text{GeV} }
\newcommand{\TEV}{ \text{TeV} }
\newcommand{\1}{\mbox{1}\hspace{-0.25em}\mbox{l}}
\newcommand{\headline}[1]{\noindent{\bf #1}}
\def\diag{\mathop\text{diag}\nolimits}
\def\Spin{\mathop\text{Spin}}
\def\SO{\mathop\text{SO}}
\def\O{\mathop\text{O}}
\def\SU{\mathop\text{SU}}
\def\U{\mathop\text{U}}
\def\Sp{\mathop\text{Sp}}
\def\SL{\mathop\text{SL}}
\def\tr{\mathop\text{tr}}
\def\mpl{M_\text{Pl}}

\def\dd{\mathrm{d}}
\def\ff{\mathrm{f}}
\def\BH{\text{BH}}
\def\inf{\text{inf}}
\def\ev{\text{evap}}
\def\eq{\text{eq}}
\def\SM{\text{sm}}
\def\Mpl{M_\text{Pl}}
\def\GeV{\text{GeV}}

\begin{titlepage}
\begin{center}

\hfill UCB--PTH--15/11\\
\hfill IPMU15--0186 \\
\hfill TTP15--037\\
\hfill KEK--TH--1859\\
\hfill October, 2015

\vspace{1.cm}
{\Large \bf 
ATLAS on-$Z$ Excess via gluino-Higgsino-singlino\\
\vspace{.3 cm}
decay chains in the NMSSM 
}

\vspace{1.0cm}
{\fontsize{14.pt}{0pt}\selectfont{
{\bf Keisuke Harigaya}$^{(a,b,c)}$,
{\bf Masahiro Ibe}$^{(c,d)}$,\\ 
\vspace{.2cm}
and
{\bf Teppei Kitahara}$^{(e,f,g)}$
}}

\vspace{1.0cm}
{\it
$^{(a)}${Department of Physics, University of California, Berkeley, California 94720, USA}\\
$^{(b)}${Theoretical Physics Group, Lawrence Berkeley National Laboratory, Berkeley, California 94720, USA}\\
$^{(c)}${ICRR, University of Tokyo, Kashiwa, Chiba 277-8582, Japan}\\
$^{(d)}${Kavli IPMU (WPI), UTIAS, University of Tokyo, Kashiwa, Chiba 277-8583, Japan}\\
$^{(e)}${Institute for Theoretical Particle Physics (TTP), Karlsruhe Institute of Technology, Engesserstra{\ss}e 7, D-76128 Karlsruhe, Germany}\\
$^{(f)}${Institute for Nuclear Physics (IKP), Karlsruhe Institute of Technology, Hermann-von-Helmholtz-Platz 1, D-76344 Eggenstein-Leopoldshafen, Germany}\\
$^{(g)}${KEK Theory Center, IPNS, KEK, Tsukuba, Ibaraki 305-0801, Japan}
}

\vspace{1.0cm}
\abstract{
Recently the ATLAS experiment has reported 3.0 $\sigma$ excess in an on-$Z$ signal region in searches for supersymmetric particles.
We find that the next-to-minimal supersymmetric standard model can explain this excess by the production of gluinos which mainly decay via  $\tilde{g} \to g \tilde{\chi}^0_{2,3} \to g Z  \tilde{\chi}^0_{1}$
 where $\tilde{\chi}^0_{2,3}$ and $\tilde{\chi}^0_1$ are the Higgsino and the singlino-like neutralinos, respectively.
We  show that the observed dark matter density is explained by the thermal relic density of the singlino-like neutralino, simultaneously.
We also discuss the searches for the Higgs sector of this scenario at the Large Hadron Collider.
}
\end{center}
\end{titlepage}
\setcounter{footnote}{0}
\renewcommand{\thefootnote}{\#\arabic{footnote}}
\setcounter{page}{1}

\hrule
\tableofcontents
\vskip .2in
\hrule
\vskip .4in

\section{Introduction}

The supersymmetric (SUSY) models are attractive candidates for physics beyond the standard model (SM)  because the hierarchy problem can be solved and  dark matter is naturally introduced when the $R$-parity is conserved.
Among various possibilities, the minimal SUSY standard model (MSSM) has been the prime candidate for the realistic supersymmetric model.
A drawback of the MSSM is, however, that it contains a dimensionful parameter $\mu$, the mass term of the Higgs multiplets. 
This reintroduces an additional fine-tuning problem, the so-called $\mu$ problem~\cite{Kim:1983dt}.
The size of $\mu$ is required to be of the order of other soft SUSY breaking parameters for successful electroweak symmetry breaking whereas they are essentially unrelated with each other.

The simplest solution of the $\mu$ problem is to introduce an additional gauge-singlet superfield  $\hat{S}$ \cite{Fayet:1974pd}
whose vacuum expectation value (VEV) is controlled by soft SUSY breaking parameters. 
By making an effective $\mu$ term generated by the VEV of $\hat S$, the size of $\mu$ is naturally interrelated to the size of the soft SUSY breaking parameters.
The next-to-minimal SUSY standard model (NMSSM) is one of the simplest singlet extensions of the MSSM 
where a discrete $\mathbb{Z}_3$ symmetry is imposed \cite{Fayet:1974pd,Nilles:1982dy,Frere:1983ag}. 
 
Recently, the ATLAS experiment has reported excess events in the SUSY particle searches  with dileptons, jets and missing transverse energy ($E_{\textrm{T}}^{\rm miss}$) in data of 20.3 fb$^{-1}$ at $\sqrt{s} = 8$ TeV~\cite{Aad:2015wqa}. 
They have observed $29$ (16 for $ee$ and 13 for $\mu \mu$) same-flavour opposite-sign dilepton pairs whose invariant masses are in the $Z$ boson mass window, 
$81\,{\GeV} < m_{\ell \ell} < 101\,\GeV$ (``on-$Z$" signal region).
The expected number of SM background events is $10.6 \pm 3.2$ pairs.
The observed event number corresponds to excess of 3.0 $\sigma$ local significance (3.0 $\sigma$ for $ee$ and 1.7 $\sigma$ for $\mu\mu$, separately). 
In this paper, we call this excess ``ATLAS on-$Z$ excess".
This excess seems to imply an existence of a gluino whose mass is lighter than 1.2 TeV or squarks lighter than 1.4 TeV\,\cite{Barenboim:2015afa}.
The caveat is, though, that  the CMS experiment has also analyzed the dileptons$+$jets$+E_{\textrm{T}}^{\rm miss}$ final state using $\sqrt{s} = 8$ TeV data in which the kinematical cut is different from the ATLAS one, and a significant excess has not been observed in the on-$Z$ signal region \cite{Khachatryan:2015lwa}.

After the report, many scenarios in the MSSM as well as in the NMSSM have been proposed to explain the ATLAS on-$Z$ excess without conflicting with constraints from various SUSY searches including the CMS on-$Z$ result \cite{Barenboim:2015afa,Vignaroli:2015ama,Ellwanger:2015hva,Allanach:2015xga,Kobakhidze:2015dra, Cao:2015ara, Dobrescu:2015asa,Cahill-Rowley:2015cha, Lu:2015wwa,Liew:2015hsa,Cao:2015zya,Collins:2015boa,Ding:2015jya,Chiang:2015iva}.  
To have on-shell $Z$ bosons in final states while escaping other SUSY search constraints,
scenarios with a gravitino as the lightest SUSY particle  (LSP) seems to be one of the simplest possibilities.
The lightest neutralino decays into a pair of a $Z$ boson and a gravitino, while the decay of colored SUSY particles into the gravitino with a large  $E_{\textrm{T}}^{\rm miss}$  are suppressed.
Unfortunately, however, the simplified general gauge mediation model with the gravitino LSP cannot explain the ATLAS on-$Z$ excess~\cite{Ellwanger:2015hva, Allanach:2015xga}, 
where the produced $Z$ bosons are rather hard due to the lightness of the gravitino and are caught in the mesh of the SUSY searches with multi  jets\,$ +\,   E_{\textrm{T}}^{\rm miss}$.

In Ref.\,\cite{Liew:2015hsa}, it has been shown that this problem can be evaded by introducing a non-MSSM massive particle,  a goldstini $\tilde{G}'$, into which the lightest neutralino mainly decays. 
Due to the massiveness of the goldstini, the $Z$ bosons are emitted softly, and hence,
the constraints from $0$ lepton\,$+$\,multi jets\,$+\,E_{\textrm{T}}^{\rm miss}$ searches become weaker.
In this scenario, similar to the gravitino LSP scenario,  the couplings between the goldstini and the MSSM particles are suppressed.
Besides, the sfermion masses are assumed to be rather larger than the gaugino masses and the lightest neutralino is assumed to be the Higgsino-like. 
With this setup, the ATLAS on-$Z$ excess is successfully explained by the gluino production via a decay chain, $\tilde{g} \to g \tilde{\chi}^0_{1,2} \to g Z \tilde{G}'$, where $ \tilde{\chi}^0_{1,2} $ are the Higgsino-like neutralinos.

The above goldstini nature is also realized in the NMSSM with a singlino-like neutralino LSP, where the singlino is the fermionic component of the additional singlet superfield $\hat{S}$. 
Actually, some literature investigated this possibility with decay chains of $\tilde{g} \to q \bar{q} \tilde{\chi}^0_2 \to  q \bar{q} Z \tilde{\chi}^0_1$  \cite{Ellwanger:2015hva,Cao:2015ara}, and  $\tilde{q} \to q \tilde{\chi}^0_2 \to q Z \tilde{\chi}^0_1 $\cite{Cao:2015zya,Ding:2015jya}, where in both cases $\tilde{\chi}^0_2 $ and $\tilde{\chi}^0_1$ are the bino and singlino-like neutralinos, respectively.

\begin{figure}[tbp]
\centering
\includegraphics[width=0.4\linewidth, bb = 0 0 427 363]{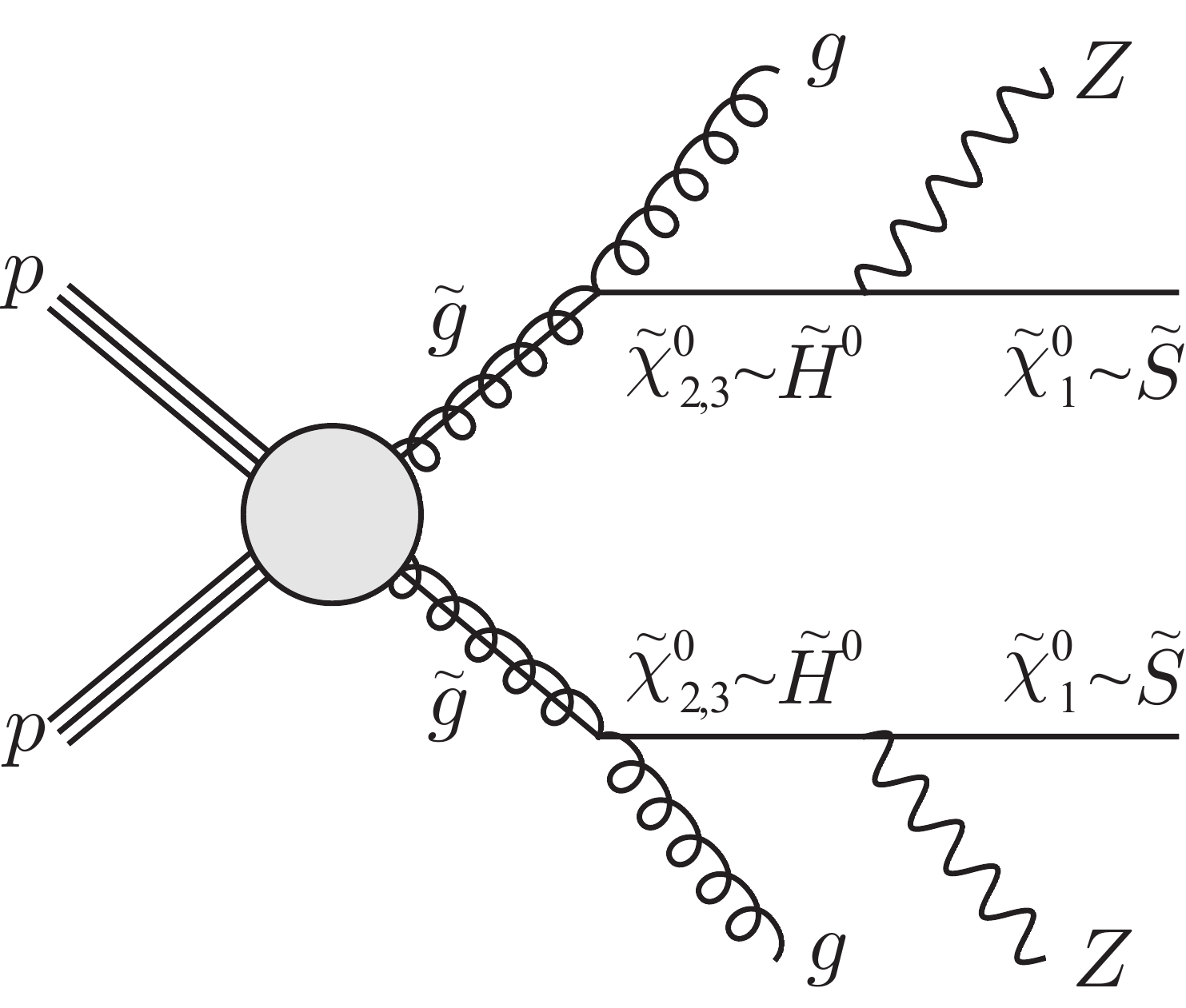}
\caption{\sl 
A typical diagram of the gluino decay in the scenario which we considered.
See section\,\ref{sec:main} for details.}
\label{fig:gluino_decay}
\end{figure}

In this paper, we discuss  another possibility in the NMSSM and consider the gluino pair production whose decay chain is $\tilde{g} \to g \tilde{\chi}^0_{2,3} \to g Z  \tilde{\chi}^0_{1}$, 
where $\tilde{\chi}^0_{2,3}$ and $\tilde{\chi}^0_1$ are the Higgsino-like and the singlino-like neutralinos, respectively (see Figure~\ref{fig:gluino_decay}).
It should be noted that the two-body gluino decay modes $\tilde{g} \to g \tilde{\chi}^0_{2,3} $ at one-loop level are the dominant ones
when the mass differences between the gluino and the Higgsinos are moderate and the squark masses are in the several TeV range~\cite{Baer:1990sc,Gambino:2005eh,Lu:2015wwa}.
Besides, as we will show, the radiative decay of the gluino reproduces the distribution of the jet multiplicity observed in the ATLAS results well.

We also investigate the  properties of dark matter in the NMSSM, and find two distinct benchmark parameter sets.
Eventually, we show that the ATLAS on-$Z$ excess and the observed relic abundance of  dark matter can be simultaneously explained in the NMSSM, without conflicting with other experimental constraints, including the CMS on-$Z$ result.

We organize the rest of this paper as follows.
In Section \ref{sec:model}, we briefly review the mass matrices of the Higgs-sector and neutralinos in the NMSSM.
Next in Section~\ref{sec:main}, we give an explanation of the ATLAS on-$Z$ excess using the NMSSM with the singlino-like neutralino LSP.
In Section~\ref{sec:DM}, the property of the dark matter around our benchmark points is discussed.
In Section~\ref{sec:higgses}, we discuss the searches for the Higgs sector at the LHC.
The final section is devoted to conclusion.

\section{NMSSM}
\label{sec:model}

In this paper, we investigate a possible explanation of the ATLAS on-$Z$ excess in the NMSSM.
Motivated by the study about the goldstini  \cite{Liew:2015hsa}, we consider the case of the singlino-like neutralino LSP.

In this section, we quickly review the mass spectrum  of the NMSSM by paying particular attention to the dependence on parameters.
The detail of the NMSSM is given in  Ref.~\cite{Ellwanger:2009dp}.
The superpotential and the scalar potential of the Higgs-sector in the NMSSM are given as 
\beq
W &=& \lambda \hat{S} \hat{H_2}  \hat{H_1} + \frac{\kappa}{3}\hat{S}^3,\label{NMSSMSP}\\ 
V &=&  (|\lambda S|^2 + m_{1}^2)|H_1|^2  + (|\lambda S|^2 + m_{2}^2)|H_2|^2  \non
& &+ \frac{g^2}{2}| ( H_1^{\dagger} H_2) |^2+ \frac{g^2 + g^{\prime 2}}{8}(|H_1|^2 - |H_2|^2)^2 \non
& &  + |\lambda H_2  H_1 + \kappa S^2 |^2+ m_{s}^2 |S|^2   + ( \lambda A_{\lambda} S H_2  H_1 + \frac{\kappa}{3} A_{\kappa} S^3 +\text{H.c.}), \label{HiggsP1}
\eeq
where fields with hats are superfields, and $\hat S$, ${\hat H}_{1,2}$  are the singlet and the  down-type and up-type Higgs doublets, respectively.
$\kappa$ and $\lambda$ are  coupling constants in the superpotential,
and $g$ and $g'$ are gauge coupling constants of the $SU(2)_L$ and $U(1)_Y$ gauge interactions.
Due to the discrete $\mathbb{Z}_3$ symmetry under which $\hat S$ as well as ${\hat H}_{1,2}$ rotate with unit charges, the superpotential does not have dimensionful parameters.
The dimensionful parameters in the scalar potential are the soft SUSY breaking parameters, i.e., $m_{1,2,s}^2$ and $A_{\lambda,\kappa}$.

For simplicity, we assume all parameters to be real in this paper.
When the electroweak symmetry is broken, three neutral scalar bosons obtain VEVs and they are expanded  around their VEVs as follows, 
\beq
H_1^0 = v_1 + \frac{H_{1R}+i H_{1I}}{\sqrt{2}}, \, ~~\, H_2^0 = v_2 + \frac{H_{2R}+i H_{2I}}{\sqrt{2}}, \, ~~\, S = v_s + \frac{S_{R} +i S_{I}}{\sqrt{2}}, \label{vev1}
\eeq
where $v^2 = v_1^2 + v_2^2  \simeq (174.1~\GeV)^2$ and we use $\tb \equiv v_2 / v_1$ in the following.
Then the effective $\mu$ term is generated, 
\beq
\mu_{\text{eff}}= \lambda v_s. \label{eff}
\eeq
Since the singlet VEV $v_s$ is the same scale as the SUSY breaking dimensionful terms, the $\mu$ problem is solved.

In the NMSSM, the singlino is the one of the neutralino components.
The neutralino mass matrix for the basis $\psi ^{0 } = \left( \tilde{B},~ \tilde{W}^0,~\tilde{H}^0_1,~\tilde{H}_2^0,~\tilde{S} \right) $ is given as
\beq
\mathcal{L}= -\frac{1}{2} (\psi ^{0})^{T} M_{\tilde{N}} \psi ^{0 } + \text{H.c.}, \label{Ne1}
\eeq
with
\beq
M_{\tilde{N} } = \begin{pmatrix} M_1 &0 &-\frac{g^{\prime} v_1}{\sqrt{2}} &\frac{g^{\prime} v_2}{\sqrt{2}}   &0 \\
0 &M_2 &\frac{g v_1}{\sqrt{2}}   &-\frac{g v_2}{\sqrt{2}}   &0 \\
-\frac{g^{\prime} v_1}{\sqrt{2}}   &\frac{g v_1}{\sqrt{2}}  &0 &-\mu_{\textrm{eff}} &-\lambda v_2 \\
\frac{g^{\prime} v_2}{\sqrt{2}}  &-\frac{g v_2}{\sqrt{2}}   &-\mu_{\textrm{eff}} &0 &-\lambda v_1 \\
0 &0 &-\lambda v_2 &-\lambda v_1 &2 \kappa v_s \end{pmatrix}. \label{Ne2}
\eeq
Typically, the mass of the singlino-like neutralino is given as $m_{\tilde{S}} \simeq | 2 \kappa v_s |$.

The Higgs mass matrices for the basis ($H',~h',~S_R$) and ($A',~S_I,~G$) (defined below) are given as 
\begin{eqnarray}
V &=&\frac{1}{2} \left( \begin{array}{ccc} H'&  h'&  S_{R} \end{array}\right) \begin{pmatrix} \tilde{\mathcal{M}}_{R 11}^2 &\tilde{\mathcal{M}}_{R 12}^2 &\tilde{\mathcal{M}}_{R 13}^2 \\ \tilde{\mathcal{M}}_{R 12}^2 &\tilde{\mathcal{M}}_{R 22}^2 &\tilde{\mathcal{M}}_{R 23}^2\\ \tilde{\mathcal{M}}_{R 13}^2 &\tilde{\mathcal{M}}_{R 23}^2 &\tilde{\mathcal{M}}_{R 33}^2\end{pmatrix}
 \left( \begin{array}{ccc} H' \\ h' \\ S_{R}\end{array} \right)  \nonumber \\ 
& & + \frac{1}{2} \left( \begin{array}{ccc} A'&  S_{I}&  G\end{array} \right)  \begin{pmatrix} \tilde{\mathcal{M}}_{I 11}^2 &\tilde{\mathcal{M}}_{I 12}^2 & 0\\ \tilde{\mathcal{M}}_{I 12}^2 &\tilde{\mathcal{M}}_{I 22}^2 & 0\\ 0 & 0& 0 \end{pmatrix}
 \left( \begin{array}{ccc}A' \\ S_{I} \\ G\end{array}\right).
\eeq
with
\beq
\tilde{\mathcal{M}}_{R 11}^2 &=&M_A^2  +M_Z^2 \sin ^2 2\beta +\lambda ^2 v^2 \cos ^2 2\beta ,\label{even11} \\
\tilde{\mathcal{M}}_{R 22}^2 &=&M_Z^2 \cos ^2 2\beta +\lambda ^2 v^2 \sin ^2 2\beta,\label{even22} \\
\tilde{\mathcal{M}}_{R 33}^2 &=&\frac{\lambda A_{\lambda}v^2}{2 v_s} \sin 2\beta + \kappa v_s (A_{\kappa} + 4 \kappa v_s) ,\\
\tilde{\mathcal{M}}_{R 12}^2 &=&M_Z^2 \sin 2 \beta \cos 2 \beta -  \lambda ^2 v^2 \sin  2\beta \cos 2 \beta ,\\
\tilde{\mathcal{M}}_{R 13}^2 &=& \lambda v (2\kappa v_s +  A_{\lambda }) \cos 2 \beta ,\\
\tilde{\mathcal{M}}_{R 23}^2 &=&\lambda v \left( 2 \lambda v_s - (2 \kappa v_s + A_{\lambda})\sin 2 \beta \right), \label{R23}\\
 \non 
\tilde{\mathcal{M}}_{I 11}^2 &=&M_A^2+\lambda ^2 v^2,\label{odd1} \\
\tilde{\mathcal{M}}_{I 22}^2 &=&\frac{\lambda A_{\lambda}v^2 }{2 v_s} \sin 2 \beta + 2 \lambda \kappa v^2 \sin 2 \beta  -3\kappa A_{\kappa} v_s ,\\
\tilde{\mathcal{M}}_{I 12}^2 &=& \lambda v  (-2\kappa v_s + A_{\lambda }), \label{odd2} 
\eeq 
where
\beq
M_{A}^2 &\equiv 
     & \frac{2 \mu_{\text{eff}} (A_{\lambda} + \kappa v_s)}{\sin 2 \beta} - \lambda ^2 v^2. \label{M_A}
\eeq
Here,  the convenient basis ($H',~h',~S_R$) and ($A',~S_I,~G$) are defined by 
\beq
 \left( \begin{array}{ccc} H_{1R} \\ H_{2R} \\ S_R \end{array} \right) &\equiv& \left( \begin{array}{ccc} \sin \beta &  \cos \beta & 0 \\ -\cos \beta & \sin \beta & 0 \\ 0 & 0 & 1\end{array} \right)  \left( \begin{array}{ccc} H' \\ h' \\ S_R\end{array} \right) ,\label{sm-like} \\
\left( \begin{array}{ccc} H_{1I} \\ H_{2I} \\ S_I\end{array} \right) &\equiv & 
 \left( \begin{array}{ccc} \sin \beta & 0  &- \cos \beta \\ \cos \beta &  0& \sin \beta \\ 0 & 1 & 0\end{array} \right) \left( \begin{array}{ccc} A' \\ S_{I} \\ G\end{array} \right).
\eeq

One of the NMSSM specific $A$ terms $A_{\lambda} $ plays two important roles in determining the properties of the singlet particles and the Higgs mass spectrum.
First, the mass scale of the heavy Higgs in Eq.~(\ref{M_A}) can be controlled by $A_{\lambda}$.
Second, a singlet-doublet mixing for the SM-like Higgs boson can also be controlled by  $A_{\lambda}$.
Even if the mass of the singlet-like Higgs boson is heavy, the effect of the singlet-doublet  mixing does not decouple.
When the lightest CP even Higgs boson is the SM-like one and $M_A$ is very large, the contribution of an off-diagonal mass term $\tilde{\mathcal{M}}_{R 23}^2$ (Eq.~(\ref{R23}))  to the mass of the SM-like Higgs boson is roughly estimated as follows,\footnote
{On the other hand, the contribution to the singlet component of the SM-like Higgs mass eigenstate is decoupled in proportion to $\tilde{\mathcal{M}}_{R 23}^2 / \tilde{\mathcal{M}}_{R 33}^2 $. }
\beq
\label{eq:higgs mass negative}
\Delta m_{h_1}^2 \sim - \frac{(\tilde{\mathcal{M}}_{R 23}^2)^2}{\tilde{\mathcal{M}}_{R 33}^2} \sim -\mathcal{O} (v^2),\label{mixing}
\eeq
which leads to a too light SM-like Higgs boson mass.
This undesirable  negative contribution via the singlet-doublet mixing  can be avoided if $\lambda \ll 1$ or when the following condition is satisfied,  
\beq
A_{\lambda} = \frac{2 \lambda v_s}{\sin 2 \beta} - 2 \kappa v_s \sim \frac{2 m_{\tilde{H}}}{\sin 2 \beta} - m_{\tilde{S}}.
\label{Alambda}
\eeq
These conditions also suppress the off-diagonal mass term $\tilde{\mathcal{M}}_{R 23}^2$.

\section{Explanation of the ATLAS on-$Z$ excess}
\label{sec:main}

In this section, we give an explanation of the ATLAS $Z+$jets$+E_{\textrm{T}}^{\rm miss}$ excess in the NMSSM.
In order to explain the signals of the ATLAS on-$Z$ excess by the gluino pair production,
we consider the following mass spectrum,
\beq
m_{\tilde{g}} &\lesssim &1~\TeV, \non
m_{\textrm{Higgsino,NLSP}} &\gtrsim& m_{\tilde{g}}  -  300~\GeV, \non
 m_{\textrm{singlino,LSP}} &\simeq& m_{\textrm{Higgsino,NLSP}}  - 100~\GeV.
 \label{massspectrum}
\eeq
The reason to choose the above mass spectrum is described below.

In order to explain the ATLAS on-$Z$ excess by the gluino production, the following items are required.
\begin{itemize}
\item
The gluino mass is required to be lighter than about $1.2$ TeV
for a large enough production cross section to explain  the number of the excess  \cite{Barenboim:2015afa}.

\item 
In the dominant gluino decay chain, at least one $Z$ boson emission like $\tilde{g} \to X + $ NLSP $ \to X + Z + $ LSP  is required, where $X$ denotes SM colored particles.

\item
Since $Z$ bosons dominantly decay hadronically, multi jets$+E_{\textrm{T}}^{\rm miss}$ searches put severe constraints.
In order to ameliorate these constraints, the $Z$ boson is required to be emitted rather softly. 
In addition, the decay of the NLSP into the Higgs boson\,$+$\,LSP is required to be suppressed so that the $Z$ boson production is enhanced.
This requirement also helps to evade the multi-jets constraints caused by the Higgs boson decay.

\item
In order to shift the jet multiplicity distribution to smaller values as is favored by the ATLAS result, the loop-induced gluino two-body decay is required to become dominant channel.
This requirement is also advantageous to evade the multi-jets constraints.

\end{itemize}

To satisfy the above conditions, we consider the following mass spectrum in the NMSSM.
First, in order to enhance the soft $Z$ boson production, we assume a slight degeneracy between the  NLSP and the LSP; 
$ m_Z <  m_{\textrm{NLSP} } - m_{\textrm{LSP} } < m_h$.

Next, in order for the gluino two-body decay to be a dominant channel, we assume that the gluino is lighter than the squarks, the bino, and the wino, while it is heavier than the Higgsinos and the singlino.%
\footnote{
This spectrum requires non-universal gaugino masses.
We note that the doublet-triplet splitting of the Higgs is naturally achieved in grand unified theories with product gauge groups~\cite{Izawa:1997he} (see Ref.~\cite{Harigaya:2015zea} and references therein for related discussions),
which in general predict non-universal gaugino masses.
}
Besides, to suppress the decay of the gluino into $t \bar{t} (t\bar{b},\,b\bar{t})$\,$+$\,Higgsino  kinematically, we assume a slight degeneracy  between the gluino and the Higgsino; 
$ m_{\tilde{g} } - m_{\textrm{Higgsino} } \lesssim 300$\,GeV.
Finally, to suppress the decay  of  the gluino into $b \bar{b}$\,$+$\,Higgsino, we take the squark masses to be of order of $10$\,TeV.
Note that the loop-induced gluino two-body decay into the gluon $+$ Higgsino is relatively enhanced by a factor $(\ln (m_{\tilde{t}} / m_t ))^2$  \cite{Baer:1990sc,Gambino:2005eh,Lu:2015wwa}.

With the above mass spectrum, the $Z$ boson is produced by the decay of the NLSP Higgsino-like neutralino into the LSP singlino-like neutralino.
In terms of the model parameters, the Higgsino mass is typically given by $|\lambda v_s|$ and the singlino mass is given by $|2 \kappa v_s|$.
Hence, $\lambda \sim 2 \kappa $ to achieve the above mass spectrum.

Altogether, we consider the mass spectrum in Eq.\,(\ref{massspectrum}), in which the typical decay chain of the gluino is $\tilde{g} \to g \tilde{\chi}^0_{2,3} \to g Z  \tilde{\chi}^0_{1}$, where $\tilde{\chi}^0_{2,3}$ and $\tilde{\chi}^0_1$ are the Higgsino and the singlino-like neutralinos, respectively.
The diagram is drawn in Figure~\ref{fig:gluino_decay}.
Note that due to small couplings between the singlino and the colored sector, the direct decay of the gluino into the singlino-like neutralino is suppressed.
This is the advantage of the singlino-like LSP compared with the bino LSP scenario where the gluino decay mode into the bino LSP with jets can be comparable to the mode into the Higgsino NLSP depending on $\tan\beta$.

Keeping above arguments in mind, we find two distinct valid parameter regions where the ATLAS on-$Z$ excess can be explained.
One is a region with a small $\lambda$ and the other is the one with a large $\lambda$.
As we will discuss in the next section, these two regions have different dark matter properties.
It should be noted that the NMSSM in the small $\lambda$ region is almost the same with the MSSM plus an additional gauge singlet fermion whose couplings to the MSSM sector are suppressed as in the simplified goldstini model \cite{Liew:2015hsa}.
In this region, the analyses of Ref.~\cite{Liew:2015hsa} would be applied.
In Table~\ref{table:point}, we show the  two benchmark parameter sets in the two distinct parameter regions
which exemplify the small $\lambda$ and the large $\lambda$ regions, respectively.
In our analysis, we use the spectrum calculator {\tt NMSSMTools 4.7.0} \cite{Ellwanger:2004xm, Ellwanger:2005dv} and also use the decay width calculator {\tt NMSDECAY} \cite{Das:2011dg}.
In the table, we also show the properties of the dark matter which are calculated using {\tt MicrOMEGAs4.1.8} \cite{Giomataris:1995fq, Belanger:2014vza} (see next section).
Note that we take $M_1 = M_2 = 1.5$ TeV in both benchmark points, and the bino and winos are decoupled enough. 

At the small $\lambda$ benchmark point, the mass of the SM-like Higgs boson is given by the radiative correction from the stop loop diagrams, which requires a large $\tb$.
This large $\tb$ leads to a large decay width of $\tilde{g} \to b \bar{b} (t\bar{b},\,b\bar{t})~ +$ Higgsino via an enhancement of the bottom Yukawa.
Even for a rather large $\tan\beta$, however, we have confirmed that the dominant decay channel of the gluino is $\tilde{g} \to g~ +$ Higgsino (see Table\,\ref{table:point}).

At the large $\lambda$ benchmark point, the mass of the SM-like Higgs boson is given by  both the  radiative correction and an additional $F$-term contribution which requires a small $\tb$ (see Eq.~(\ref{even22})).
Such a large $\lambda$ also brings the undesirable negative contribution to the lightest CP-even Higgs mass in Eq.~(\ref{mixing}).
To avoid this problem, we choose $A_{\lambda}$ that suppresses this negative contribution according to Eq.~(\ref{Alambda}).%
\footnote{The singlet component of the lightest CP-even Higgs boson is suppressed by a large singlet scalar mass.
Then the Higgs couplings are almost equivalent  to the ones in the SM even if we do not choose such a  $A_{\lambda}$.}
Note that when one takes $A_{\lambda}$ to be $\mathcal{O}(1)$\,TeV, this equation also suggests the small $\tb$.
 The large $\lambda$ also leads to a certain singlino-Higgsino mixing, so that the decay branch $\tilde{g} \to g~ +$ singlino-like neutralino exists. 
Even with such contributions, we have again confirmed that the dominant decay channel is  $\tilde{g} \to g~ +$ Higgsino (see Table\,\ref{table:point}).

To investigate the number of the SUSY events in the ATLAS on-$Z$ search and to check the experimental constraints from the other SUSY searches at the ATLAS and CMS collaborations, we use  {\tt CheckMATE 1.2.1} \cite{Drees:2013wra, Cao:2015ara} which incorporates {\tt DELPHES\,3}~\cite{deFavereau:2013fsa} and {\tt FastJet}\,\cite{Cacciari:2006sm} internally.
Signal events are generated by {\tt MadGraph5  v2.2.3}\,\cite{Alwall:2011uj} connected to {\tt Pythia 6.4}\,\cite{Sjostrand:2006za}
where the MLM matching scheme is used with a matching scale at $150$\,GeV~\cite{Alwall:2007fs}.
The parton distribution functions are {\tt CTEQ6L1}\,\cite{Pumplin:2002vw}.
We use the gluino production cross sections at the next-to-leading-logarithmic accuracy given in Ref.~\cite{NLL} with {\tt NLL-fast}~\cite{Beenakker:2015rna,Beenakker:1996ch,Kulesza:2008jb,Beenakker:2009ha}.

\begin{table}[tbp]
\caption{\sl 
Two benchmark parameter sets. In the line of Br($\tilde{g} \to $ others), parentheses represent the dominant decay channel.
In both benchmark points, the lightest neutralino $\tilde\chi_1^0$, the second lightest CP-even Higgs $h_2$, and the lightest CP-odd Higgs $a_1$ 
are the mass eigenstates dominated by singlet contributions. 
}
\begin{center}
\label{table:point}
\begin{tabular}{|c | c c|}
\hline \hline
 & small $\lambda$ & large $\lambda$
  \\ \hline\hline
$\lambda$  & 0.080   & 0.435
 \\ 
$\kappa$ & 0.033  & 0.185
 \\ 
 $\tb$ & 30   & 3
 \\ 
 $A_{\lambda}$ [\GeV] & $- 146$   & 1500
 \\ 
 $A_{\kappa} $ [\GeV] & $- 50$  & $- 200$ 
 \\ 
 $\mu_{\textrm{eff}}$ [\GeV] & 620  & 600  
 \\ 
 $m_{\tilde{q}} $ [\TeV] & 10   & 10
 \\ 
 $A_{q}$ [\TeV] & 6   & 0
 \\ 
 \hline \hline
 $m_{\tilde{g}}$ [\GeV]  & 900   & 925
 \\ 
 $m_{\tilde{\chi}^0_{3}}$  [\GeV]   & 641  & 625
 \\ 
  $m_{\tilde{\chi}^0_{2}}$   [\GeV]   & 636  & 624
 \\ 
   $m_{\tilde{\chi}^0_{1}}$  [\GeV]    & 527   & 519
 \\ 
    $m_{\tilde{\chi}^{\pm}_{1}}$  [\GeV]    & 637   & 615
 \\ 
   $m_{h_1}$   [\GeV]   & 125   &  125
 \\ 
   $m_{h_2}$ [\GeV]    & 500   & 453
 \\ 
  $m_{h_3}$  [\GeV]   & 1061   & 1831
 \\ 
   $m_{a_1}$   [\GeV]   & 195   & 439
 \\ 
   $m_{a_2}$  [\GeV]    & 1061  & 1829
 \\ 
    $m_{H^{+}}$  [\GeV]    &  1056   & 1822
 \\ 
 \hline \hline
   Br($\tilde{g} \to g \tilde{\chi}^0_{2,3}$)   & 0.79   & 0.70
 \\ 
Br($\tilde{g} \to g \tilde{\chi}^0_{1}$)    & 0.008  & 0.12
 \\ 
  Br($\tilde{g} \to $ others)   & 0.20 ($b\bar{b}\tilde{\chi}^0_{2,3}$)   & 0.18 ($tb\tilde{\chi}^{\pm}_1$)
 \\ 
   Br($\tilde{\chi}^0_{2,3}  \to Z \tilde{\chi}^0_{1}$  )   & 1.00   & 1.00
 \\ 
 \hline \hline
   SUSY events in ATLAS on-$Z$ & 14  & 14
 \\ 
   $\Omega_{\tilde{\chi}} h^2$  & 0.118  & 0.121
 \\ 
   $\sigma_{\textrm{SI}}$ [cm$^2$] & $4.0 \times 10^{-47}$   & 2.8 $ \times 10^{-45}$
 \\ 
Higgs coupling $\kappa_V$ & 0.997  & 0.9997
 \\ 
 Higgs coupling $\kappa_b$ &  1.02  & 1.02
 \\
\hline \hline
\end{tabular}
\end{center}
\end{table}

As a result, we find that the SUSY events can explain the ATLAS on-$Z$ excess within $1\,\sigma$ without conflicting with any LHC constraints at 95\%\,CL, including the CMS on-$Z$ result for  both benchmark points.
In addition,  the lightest CP-even Higgs boson mass is $125\,\GeV$\footnote{In the {\tt NMSSMTools 4.7.0}, the Higgs boson masses are calculated at a full one-loop plus  a  two-loop $\mathcal{O}(\alpha_t \alpha_s + \alpha_b \alpha_s ) $ level  \cite{Degrassi:2009yq}.
When the squark mass is taken to be $10\,\TeV$, a theoretical uncertainty of the SM-like Higgs boson mass from such the corrections is about $5\,\GeV$ in the MSSM case.
Once a resummation of large logs  is included as higher-loop corrections, this uncertainty can be reduced to around $1\,\GeV$ and the central value is raised about $5\,\GeV$ (see e.g. Ref.~\cite{Kitahara:2015wia}).
} 
and the relic abundance of the singlino-like neutralino is also consistent with the observed values within $2\,\sigma$ \cite{Ade:2015xua}. 

Before closing this section, let us comment on the distributions of the jet multiplicity, $E_{\rm T}^{\rm miss}$ and $H_T$ in the on-$Z$ signal region.
In Fig.\,\ref{fig:jet},
we show the distributions of the jet multiplicity, $E_{\rm T}^{\rm miss}$ and $H_T$ for the small $\lambda$ and large $\lambda$ benchmark points, respectively.
Note that we combine the $ee$ and $\mu \mu$ channels of the ATLAS results, although the efficiencies of the channels are different from each other.
As shown in Ref.~\cite{Aad:2015wqa}, the jet multiplicity of the observed data is typically $2$--$5$ jets while the multiplicity larger than $6$ is disfavored. 
For both benchmark points, the jet multiplicity of the NMSSM contributions is peaked at around $4$--$5$, which originates from two gluon jets in two gluino decays,  two quarks from a $Z$-boson decay in one of the gluino decay chain, and an occasionally radiated jet.
The predicted distributions of the jet multiplicity fit the ATLAS data very well, which confirms the advantage of the dominance of the gluino two body decay \cite{Liew:2015hsa}.
It can be seen that the distributions of $E_{\rm T}^{\rm miss}$ and $H_T$ are also consistent with the data.

\begin{figure}[tbp]
\begin{center}
\includegraphics[width=0.55\linewidth, bb = 0 0 313 207]{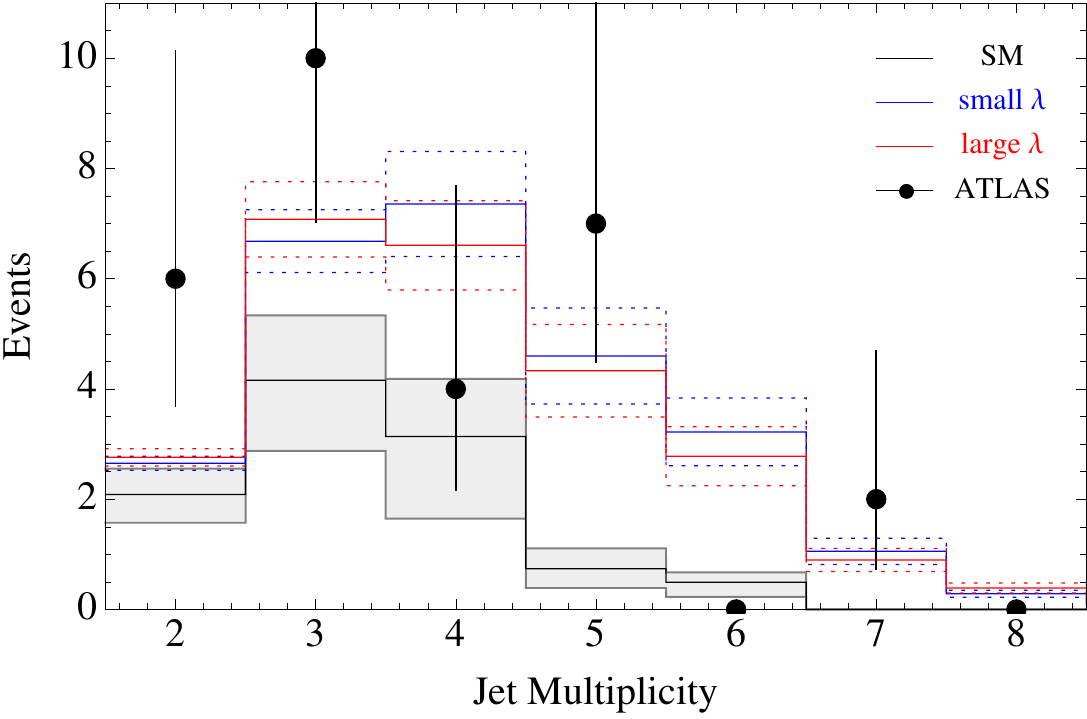}\\
\vspace{.2cm}
\includegraphics[width=0.57\linewidth, bb = 0 0 313 209]{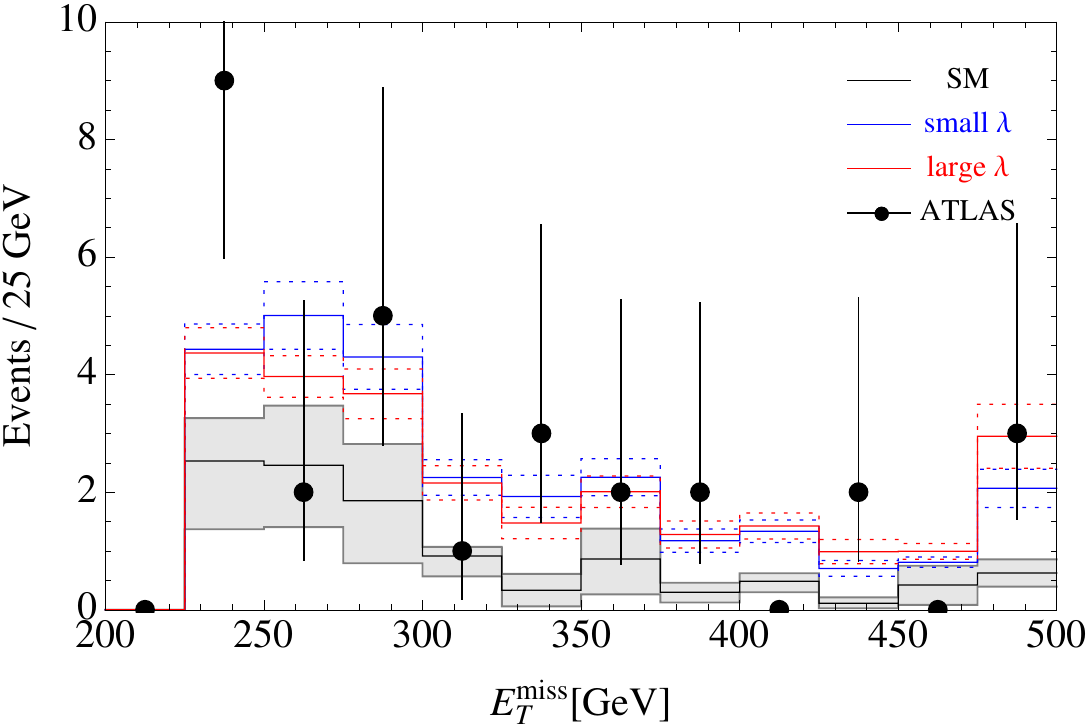}\\
\vspace{.25cm}
\includegraphics[width=0.55\linewidth, bb = 0 0 313 206]{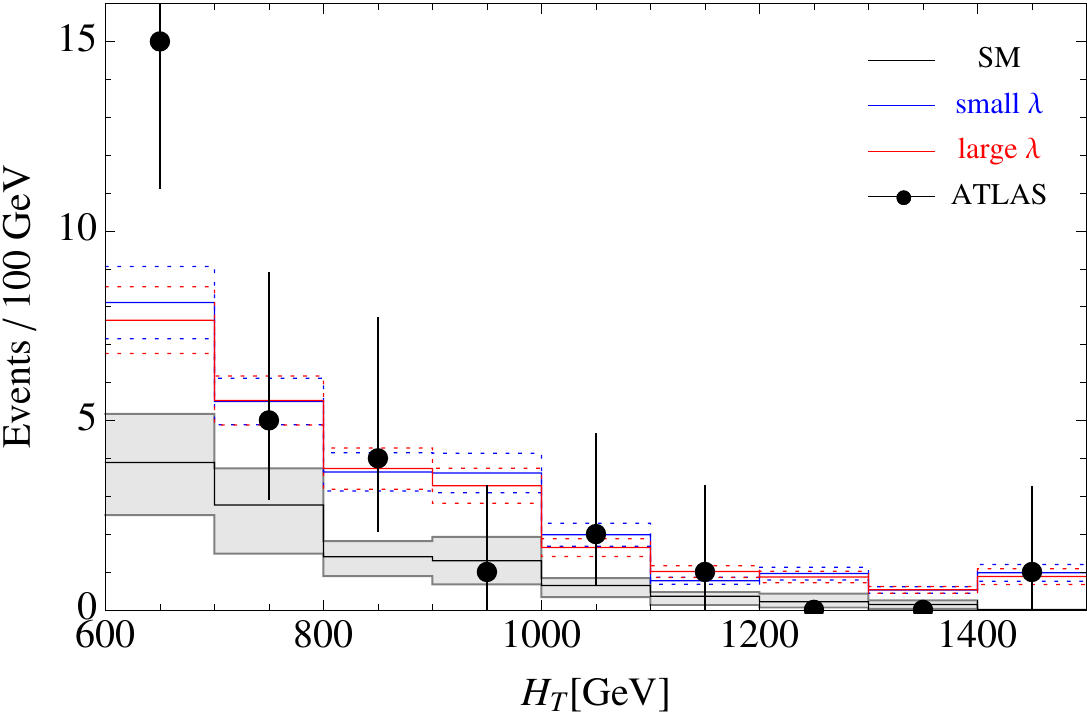}
\caption{\sl
The distributions of the jet multiplicity (top), $E_{\rm T}^{\rm miss}$ (middle) and $H_T $ (bottom) in the ATLAS on-$Z$ signal region.
The ATLAS data points are shown by black points with combined errors.
The black line shows the Standard Model  contributions whose errors are shown by gray shade.
The expected distributions for the two benchmark points are shown by the blue (small $\lambda$) and the red (large $\lambda$) lines.
The dotted lines represent the error of the NMSSM contributions which
is dominated by the uncertainty of the gluino production cross section of about $30$\%.
We ignore the uncertainty of the estimation of the efficiency in {\tt CheckMATE 1.2.1}.
}
\label{fig:jet}
\end{center}
\end{figure}

\section{Dark matter property}
\label{sec:DM}

In this section, we discuss the properties of dark matter in the small and large $\lambda$ regions as exemplified in the above two benchmark points (see Table\,\ref{table:point}).
In the small $\lambda$  region, the lightest neutralino is almost singlino-like.
In most cases, such a singlino-like LSP results in overabundant dark matter in the universe because its annihilation cross section is too small. 
The overabundance is, however, avoided  when the heavy Higgs masses are about twice of the mass of the singlino dark matter. 
Such a mass spectrum can be achieved by tuning the parameter $A_{\lambda}$ \cite{Belanger:2005kh}.
With this tuning, the singlino-like neutralino annihilates resonantly via  the $s$-channel heavy Higgs boson exchange, which significantly enhances the annihilation cross section.%
\footnote{The other possibility is a resonant annihilation  via the SM Higgs or $Z$ boson \cite{Belanger:2005kh,Cao:2012fz, Ishikawa:2014owa}.
However, due to a large mass difference between the NLSP and singlino, the emitted $Z$ boson becomes energetic.
} 
In this way, the observed dark matter density is explained by the singlino-like neutralino at the small $\lambda$ benchmark point in spite of the weakness of couplings.
It should be noted that due to a suppressed coupling between the singlino and the SM-like Higgs boson, the spin-independent dark matter-nucleon elastic scattering cross section is much lower than the reaches in currently proposed direct dark mater searches.

In the large $\lambda$ region, on the other hand, there is a certain singlino-Higgsino mixing.
The Higgsino components in the LSP can enhance the annihilation cross section, so that the tuning of the heavy Higgs mass is not required.
In addition, sizable $\kappa$ can  also enhance the annihilation cross section.
It is because that contributions of diagrams of $s$-cannel $a_s$ boson exchange into fermions \cite{Hasenkamp:2013opa} and $t$-channel $\tilde{\chi}^0_1$ exchange into $h_s + a_s$ \cite{Belanger:2005kh} become efficient when $\kappa$ is sizable, where both $a_s$ and $h_s$ are the singlet-like scalar bosons.
In fact, at the large $\lambda$ benchmark point, the contribution of a process $\tilde{\chi}^0_1 \tilde{\chi}^0_1 \to t \bar{t}$  to the annihilation cross section is $79\,\%$, and  $\tilde{\chi}^0_1 \tilde{\chi}^0_1 \to  h_2 a_1$ is $15~\%$, where a dominant decay channel of  $a_1$ is $t \bar{t}$.

The singlino-Higgsino mixing also contributes to the spin-independent dark matter-nucleon  elastic scattering cross section at the large $\lambda$ benchmark point.
We found that the typical scattering cross section around the benchmark point is a little smaller than a current experimental bound by LUX \cite{Akerib:2013tjd}, 
and proposed future experiments for the direct dark matter search can probe around this benchmark point \cite{Akerib:2012ys,Aprile:2012zx,Hiraide:2015cba,DarkSide}.

\section{Higgs sector searches at the LHC}
\label{sec:higgses}

In this section, we discuss the searches for the Higgs sector at the LHC.
First, we consider the heavy doublet-like Higgses $h_3$ and $a_2$.
Successful parameter points in the small $\lambda$ region always predict 
that the heavy doublet-like Higgses have masses around $1$\,TeV
so that the singlino-like neutralino annihilates resonantly via the s-channel  heavy Higgses exchanges.
Interestingly, the $14$\,TeV LHC with a luminosity of $300$ fb$^{-1}$  can probe such a charged Higgs directly by the channel of $pp\to tb H^{+} \to tbtb$   due to a $\tb$ enhancement of the bottom Yukawa \cite{Hajer:2015gka}.%
\footnote{
In the small $\lambda$ region, another promising  channel for the heavy doublet-like Higgs searches  is $pp \to h_3/ a_2 \to \tau \tau $ \cite{Arbey:2013jla, Li:2013nma}.} 
In the large $\lambda$ region, although the production cross section of the heavy doublet-like Higgses is small,
a high-luminosity LHC can probe them through the top Yukawa via the same channel   \cite{Hajer:2015gka}.

Next, we consider the singlet-like scalars $h_2$ and $a_1$.
The masses of the singlet-like scalars tend to be within the TeV range in our scenario. 
This is because typical masses of the singlet scalars are controlled by $\kappa v_s$, which also determines the singlino mass and is taken as about $500$\,GeV in our scenario.

The singlet-like CP-even Higgs boson $h_2$ mainly decays into $WW$ and $ZZ$ due to an enhancement of the longitudinally polarized gauge bosons when the decay width into double Higgs, $h_1 h_1$, is  suppressed.
In our benchmark points, we find that
the gluon-gluon  (vector boson) fusion cross sections of $pp \to h_2 (jj) \to WW (jj)$ are  $8$\,fb\,$(1\,$fb$)$ at the small $\lambda$ benchmark point, and $0.2$--$20$\,fb\,$(0.2$--$4$\,fb$)$ at the large $\lambda$ benchmark point for $\sqrt{s} = 14\,\TeV$, where we vary
$A_{\lambda}$ in the range of $1500$--$1100$~GeV in the large $\lambda$ benchmark point.
Within this range of $A_\lambda$, the reduction of the SM-like Higgs boson mass (see Eq.~(\ref{eq:higgs mass negative})) can be compensated by taking a large trilinear coupling between the up-type Higgs doublet and stops.
For a smaller $A_\lambda$, $h_2$ contains more doublet components and hence the production cross section is larger. 
The current experimental upper bounds for the gluon-gluon  (vector boson) fusion cross sections  are $200$ ($100$) fb for $m_{h_2} = 500\,\GeV$ at $\sqrt{s} = 8\,\TeV$ \cite{Aad:2015agg}.

On the other hand, the singlet-like CP-odd Higgs boson $a_1$ mainly decays into $t\bar{t}$.
The cross section of $pp \to a_1 \to t\bar{t}$ is  $1$\,fb   at $\sqrt{s} = 14\,\TeV$ at the large $\lambda$ benchmark point.%
\footnote{
In the small $\lambda$ benchmark point, 
$a_1 \to t\bar{t}$ is forbidden by the kinematics, and mainly decays into $b\bar{b}$. 
However, $pp\to a_1 \to b\bar{b}$ cannot be probed due to enormous background. }
The current upper bound is $2$\,pb for $m_{a_1} = 500\,\GeV$  at $\sqrt{s} = 8\,\TeV$ \cite{Aad:2015fna}.
Therefore, the direct search for such the singlet-like Higgses would be challenging.
It is recently shown that the precision measurement of the double Higgs production can probe the singlet-like Higgs bosons through the interference effect between a resonance production of the singlet-like Higgs and SM processes \cite{Dolan:2012ac, No:2013wsa, Chen:2014ask,Martin-Lozano:2015dja, Dawson:2015haa}. 
This resonance signal, however,  would become narrower than results in the literature due to the small singlet-doublet mixing in our benchmark points.

\section{Conclusion}
\label{sec:summary}

In this paper, we have studied a possible explanation of the ATLAS on-$Z$ excess in the NMSSM by the gluino production via typical decay chains, $\tilde{g} \to g \tilde{\chi}^0_{2,3} \to g Z  \tilde{\chi}^0_{1}$, with $\tilde{\chi}^0_{2,3}$ and $\tilde{\chi}^0_1$ being the Higgsino and the singlino-like neutralinos, respectively.
We found two distinct benchmark parameter sets.
At the benchmark points, the observed dark matter density is also explained by the thermal relic abundance of the singlino-like neutralino.
In addition, it is found that the expected distributions of the jet multiplicity, $E_{\rm T}^{\rm miss}$ and $H_T$ for our benchmark points are consistent with the ATLAS data.

In the small $\lambda$ region, we find that the 14\,TeV LHC  with a luminosity of $300$ fb$^{-1}$  can probe the $1$\,TeV charged Higgs directly by the channel of $pp\to tb H^{+} \to tbtb$.
On the other hand, in the large $\lambda$ region, we find that proposed future experiments for the direct dark matter search can probe our scenario at around the benchmark point.

\section*{Acknowledgements}
The authors would like to thank Satoshi Shirai for useful discussions.
This work is supported by Grants-in-Aid for Scientific Research from the Ministry of Education, Culture, Sports, Science, and Technology (MEXT)
KAKENHI, Japan,
No.~24740151 and  No.~15H05889 (M.\,I.), and No.~25105011 (M.\,I. and T.\,K.);
Grant-in-Aid No.~26287039 (M.\,I.) from the Japan Society for the Promotion of Science (JSPS) KAKENHI;
and by the World Premier International Research Center Initiative (WPI), MEXT, Japan (M.\,I.).
K.\,H.\,was supported in part by a JSPS Research Fellowship for Young Scientists.


\end{document}